\documentclass[11pt]{article}
\usepackage{epsfig}
\usepackage{amssymb}
\usepackage{amsfonts}
\usepackage{subfigure}
\def\beq{\begin{equation}}
\def\eeq{\end{equation}}
\def\bea{\begin{eqnarray}}
\def\eea{\end{eqnarray}}
\def\ksl{\hbox{\hbox{${k}$}}\kern-1.9mm{\hbox{${/}$}}}

\newcommand{\nn}{\nonumber}
\newcommand{\text}{\rm}

\newcommand{\psl}{p \! \! \!  /}
\newcommand{\tRE}{\widetilde{\rm Re}}

\def\lsim{\raise0.3ex\hbox{$\;<$\kern-0.75em\raise-1.1ex\hbox{$\sim\;$}}} 

\def\gsim{\raise0.3ex\hbox{$\;>$\kern-0.75em\raise-1.1ex\hbox{$\sim\;$}}}

\begin{document}

\begin{center}

{\bf \Large Mass Corrections to Flavor-Changing Fermion-Graviton Vertices\\
\vspace{.5cm}
 in the Standard Model} 

\vspace{1.5cm}
{\bf Claudio Corian\`{o}$^{a}$, Luigi Delle Rose$^{a}$, 
Emidio Gabrielli$^{b,c}$\footnote{
On leave of absence from Dipartimento di Fisica  Universit\`a di 
Trieste, Strada Costiera 11, I-34151 Trieste \\}, and Luca Trentadue$^{d}$}
\vspace{1cm}

{\it $^{(a)}$Dipartimento di Matematica e Fisica "Ennio De Giorgi", 
Universit\`{a} del Salento and \\ INFN-Lecce, Via Arnesano, 73100 Lecce, Italy\footnote{claudio.coriano@le.infn.it, luigi.dellerose@le.infn.it, 
emidio.gabrielli@cern.ch, luca.trentadue@cern.ch}}\\

\vspace{1cm}
{\it $^{(b)}$ NICPB, R\"avala 10, Tallinn 10143, Estonia\\
$^{(c)}$ INFN, Sezione di Trieste, Via Valerio 2, I-34127 Trieste, Italy \\}
\vspace{1cm}
{\it$^{(d)}$ Dipartimento di Fisica e Scienze della Terra "Macedonio Melloni", Universit\`a di Parma and
INFN, Sezione di Milano Bicocca, Milano, Italy
\\
}
\vspace{.5cm}
\begin{abstract} 
In a previous study, the flavor-changing fermion-graviton interactions 
have been analyzed in the framework of the standard model, where
analytical results for the relevant form factors were obtained at 
the leading order in the external fermion masses.
These interactions
arise at one-loop level by the charged electroweak corrections 
to the fermion-graviton vertex, when the off-diagonal flavor transitions in the 
corresponding charged weak currents are taken into account.
Due to the conservation of the energy-momentum tensor, the corresponding
form factors turn out to be finite and gauge invariant when
external fermions are on-shell.
Here we extend this previous analysis by including the exact dependence
on the external fermion masses. Complete analytical results are provided for all the relevant form factors to the flavor-changing fermion-graviton transitions.
\end{abstract}
\end{center}

\newpage
\section{Introduction}
In a previous analysis \cite{Coriano:2012cr}, following the study of Ref.
\cite{Degrassi:2008mw}, we have discussed the structure of the 
perturbative corrections to the graviton-fermion-antifermion 
($T f \bar{f}$) vertex in the Standard Model (SM), focusing our attention on the flavor diagonal sector. On the other hand, in  \cite{Degrassi:2008mw} 
the one-loop electroweak corrections which generate the off-diagonal 
graviton-fermion-antifermion vertex, were computed at the leading order 
in the external fermion masses. 
These studies address the structure of the interactions between the fermions of the Standard Model and gravity, beyond leading order in the weak coupling, 
which have never been presented before in their exact expressions. The choice of an external (classical) gravitational background allows to simplify 
the treatment of such interactions where the coupling is obtained by the insertion of the symmetric and improved energy-momentum tensor (EMT) into ordinary correlators of the Standard Model. 

We have addressed some of the main features of the perturbative structure of these corrections, presenting their explicit form, parameterized in terms of a certain set of form factors. We have also discussed some of their radiative properties with regard to their infrared finiteness and renormalizability, the latter being inherited directly from the Standard Model, when the coupling of the Higgs to the gravitational background is conformal. 

In general, one expects that such corrections are small,  although they could become more sizeable in theories with a low gravity scale 
 \cite{ArkaniHamed:1998rs, Antoniadis:1998ig, ArkaniHamed:1998nn, Randall:1999ee,Randall:1999vf, Dvali:2000hp}.  In particular, one can consider the possibility of including, in these constructions, backgrounds which are of dilaton type, with dilaton fields produced by metric compactifications. The same vertices characterize the interaction of a dilaton of a spontaneously broken dilatation symmetry with the ordinary fields of the Standard Model \cite{Goldberger:2007zk, Campbell:2011iw, Coriano:2012nm, Barger:2011hu}. This second possibility is particularly interesting, in view of the recent discovery of a Higgs-like scalar at the LHC. 
 
 Perturbative studies of these vertices have their specific difficulties due to the proliferation of form factors, and the results have to be secured by consistency checks using some relevant Ward identities. These need to be derived from scratch using the full Lagrangian of the Standard Model, as discussed in \cite{Coriano:2011zk} and  \cite{Coriano:2012cr}. In this study we are going to reconsider the gravitational form factor of a Standard Model fermion in the presence of a background graviton in the off-diagonal flavor case, which has been discussed before \cite{Degrassi:2008mw}, extending that analysis. One of the goals of this re-analysis is to include all the mass corrections to the related form factors, which has not been given before. 
 These corrections are important in order to proceed in a follow-up work with a systematic phenomenological study. 
 In this respect, mass corrections are important in order to extract the exact behavior of these form factors in the infrared and ultraviolet limits, 
 which may be of experimental interest. We have compared our new results against the previous ones given in \cite{Degrassi:2008mw} in the limit of massless external fermions and found complete agreement.  

Our work is organized as follows. In section 2 we give the theoretical framework of the Standard Model Lagrangian in a curved space-time, assuming as a background metric the usual 4-dimensional one. In section 3 we discuss the technical
details for the calculation of 
the flavor-changing fermion matrix elements of the energy momentum tensor at the leading order in perturbation theory. The contributions coming 
from the counterterms of the 
wave-function renormalization is separately discussed in section 4. In section
5 we analyse the role played by the Ward identity from the conservation 
of the EMT, while in section 6 we give the results for the complete set of 
the relevant form factors entering in the flavor-changing matrix elements 
of EMT. Finally, in section 7 we give our conclusions.

\section{The Lagrangian}
We follow closely the layout of our previous work  \cite{Coriano:2012cr} where more details concerning the general structure of the action describing the coupling of the Standard Model to gravity can be found.   
We just recall, in order to make our treatment self-contained, that the interaction of the Standard Model fields with gravity is described by the action integral
\bea
 S = -\frac{1}{\kappa^2} \int d^4 x \sqrt{-g} R + \int d^4 x \sqrt{-g} \mathcal L_{SM} 
\eea
together with a term of improvement 
\bea
S_I = \chi \int d^4 x \sqrt{-g} R \, H^\dag H
\eea
where $R$ is the scalar curvature and $H$ is the Higgs doublet. The identification of this second term goes back to \cite{Callan:1970ze}. The coefficient $\chi$ is  an arbitrary parameter which at the special value $\chi_c \equiv 1/6$ renders the Lagrangian conformally symmetric when the scalar is massless and guarantees its renormalizability at the leading order in the gravitational $\kappa$, where $\kappa^2 = 16 \pi G_N$, with $G_N$ being the gravitational Newton's constant. 
For instance, in the case of the Higgs field, this takes place if we drop the quadratic terms in the Higgs potential. As in our previous work, our results are given for an arbitrary $\chi$. 

The Standard Model action $S_{SM}$ is obtained by promoting the ordinary SM Lagrangian to a curved background, which is parametrized by the metric expansion $g_{\mu\nu} = \eta_{\mu\nu} + \kappa \, h_{\mu\nu}$ where $\eta_{\mu\nu} = (+,-,-,-)$ and $h_{\mu\nu}$ denotes the fluctuation of the graviton field around the flat limit. At this order the graviton-matter interactions, which we are going to evaluate in the flavor changing fermion sector, are described by Green's functions with a single insertion of the energy-momentum tensor
\bea
T_{\mu\nu} = \frac{2}{\sqrt{-g}} \frac{\delta}{\delta g^{\mu\nu}} \bigg[ S_{SM} + S_I \bigg]_{g=\eta}.
\eea
The complete Standard Model energy-momentum tensor includes several contributions which can be found in \cite{Coriano:2011zk}.

\section{The perturbative expansion}
\label{Sec.PertExp}
The interaction of one graviton with two fermions of different flavor is summarized by the vertex function
\bea
\hat T^{\mu\nu} \equiv  i \langle f_i, p_i| T^{\mu\nu}(0) | p_j, f_j\rangle
\eea
that we intend to study. Here $p_j$ ($f_j$) 
and $p_i$ ($f_i$) indicate the momenta (flavor) of initial and final 
fermions respectively. We will restrict to  the case of 
flavor-changing transitions, namely $f_i\neq f_j$.
In order to simplify the results we will also use the combinations of momenta $p = p_i + p_j$ and $q = p_j - p_i$. The external states are taken on their mass shell, $p_i^2 = m_i^2$ and $p_j^2=m_j^2$ 
and can be either leptons or quarks. From now on, we will assume that 
$m_i\neq m_j$. In the last case, since the EMT is diagonal in color space, the color structure is rather trivial and therefore we omit it.  

At tree level the flavor-changing gravitational interaction is absent so that the leading order contribution comes from the quantum corrections. 
At one loop level, instead, we decompose the $\hat T^{\mu\nu}$ matrix element as
\bea
\hat T^{\mu\nu} = \hat T^{\mu\nu}_W + \hat T^{\mu\nu}_{CT}
\eea
where the first term on the r.h.s represents the pure vertex corrections 
induced by the $W^\pm$ gauge boson and its Goldstone $\phi^\pm$ exchanges,
while the last term, $\hat T^{\mu\nu}_{CT}$, includes 
the usual counterterms (CT) coming from the wave-function renormalization 
insertions on the external legs.  
The inclusion of this last term  $\hat T^{\mu\nu}_{CT}$ is needed
in order to get finite results for the matrix element $\hat T^{\mu\nu}$, as it 
will be extensively  discussed in section \ref{Sec.Renorm}. 
The finiteness of the result is just a consequence of the
non-renormalization theorem of conserved currents, when applied to the case 
of a conserved EMT.

\begin{figure}[t]
\centering
\subfigure[]{\includegraphics[scale=0.7]{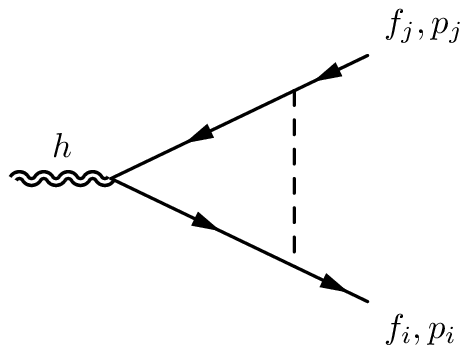}} \hspace{.5cm}
\subfigure[]{\includegraphics[scale=0.7]{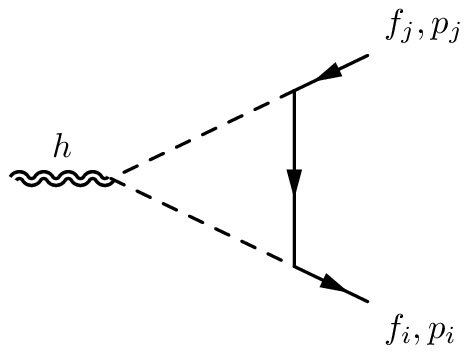}} \hspace{.5cm}
\subfigure[]{\includegraphics[scale=0.7]{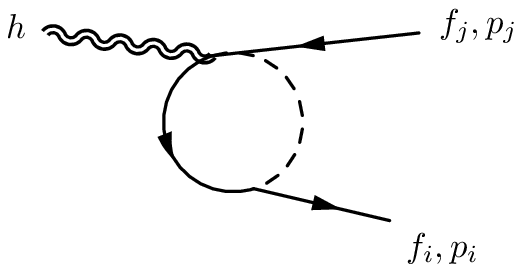}} \hspace{.5cm}
\subfigure[]{\includegraphics[scale=0.7]{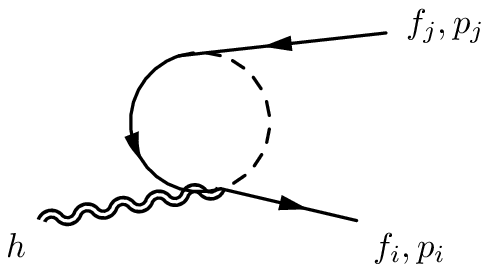}}
\caption{Diagrams of one-loop SM corrections to the flavor-changing graviton fermion vertex, where $f_{i,j}$ and $p_{i,j}$ specify the fermion flavors and corresponding momenta respectively, with $f_i \neq f_j$. \label{diagrams}}
\end{figure}

We choose to work in the $R_{\xi}$ gauge where every massive gauge field is always accompanied by its unphysical longitudinal part. The diagrammatic expansion of $\hat T^{\mu\nu}_W$ is depicted in Fig.\ref{diagrams} and is made of one contribution of triangle topology plus contact terms (see Fig. \ref{diagrams}  (c) and (d))  with a fermion and a graviton pinched on the same external point. The Feynman rules are listed in Appendix \ref{feynrules}. The computation of these diagrams has been performed in dimensional regularization using the on-shell renormalization scheme. To check the correctness of our results the Ward identity of the conservation of the EMT, which will presented in section \ref{Sec.WardId}, has been verified explicitly. 

where $f_{i,j}$ and $p_{i,j}$ specify the fermion flavors and corresponding momenta respectively,
with $f_i \neq f_j$

Due to the chiral V-A nature of the $W$ interactions, we expand the flavor changing matrix element in terms of invariant amplitudes 
$f_k$ and tensor operators $O_k$ as 
\bea
\label{hatT}
\hat T^{\mu\nu} = i \frac{G_F}{16 \pi^2 \sqrt{2}} \sum_{k=1}^{12} f_k(p,q) \, \bar u_i(p_i) O^{\mu\nu}_k u_j(p_j)
\eea
with  the tensor operators given by
\bea
\begin{array}{ll}
O_1^{\mu\nu} = \left( \gamma^{\mu} p^{\nu}+\gamma^{\nu} p^{\mu}\right) P_L \qquad \qquad
& O_7^{\mu\nu}=  \eta^{\mu\nu} \, M_- \\
O_2^{\mu\nu}= \left( \gamma^{\mu} q^{\nu}+\gamma^{\nu} q^{\mu}\right) P_L 
& O_8^{\mu\nu}=  p^{\mu}p^{\nu} \, M_-  \\
 O_3^{\mu\nu}= \eta^{\mu\nu} \, M_+  
&O_9^{\mu\nu}=  q^{\mu}q^{\nu} \, M_-  \\
O_4^{\mu\nu}= p^{\mu}p^{\nu} \, M_+
&O_{10}^{\mu\nu}= \left(p^{\mu}q^{\nu}+q^{\mu}p^{\nu}\right) M_-\\
O_5^{\mu\nu}=  q^{\mu}q^{\nu} \, M_+
& O_{11}^{\mu\nu}= \frac{m_i m_j}{m_W^2}  \left( \gamma^{\mu} p^{\nu}+\gamma^{\nu} p^{\mu}\right) P_R \\
O_6^{\mu\nu}=  \left(p^{\mu}q^{\nu}+q^{\mu}p^{\nu}\right)\, M_+
&O_{12}^{\mu\nu}= \frac{m_i m_j}{m_W^2}  \left( \gamma^{\mu} q^{\nu}+\gamma^{\nu} q^{\mu}\right) P_R 
\end{array}
\label{basis}
\eea
where $P_{L,R}=(1\mp \gamma_5)/2$ and $M_{\pm}\equiv m_j P_R\pm m_i P_L$, and 
$u_{i,j}(p_{i,j})$ are the corresponding fermion
bi-spinor amplitudes in momentum space.\\ 
This is the most general rank-2 tensor basis that can be built out of two momenta, $p$ and $q$, a metric tensor and Dirac matrices $\gamma^\mu$ and $\gamma^5$. Its expression has been given in \cite{Degrassi:2008mw}. \\
For the form factors appearing in Eq.(\ref{hatT}) we use the following notation
\bea
f_k(p, q) = \sum_f  \lambda_f  \, F_{k}(p,q,m_f) \,,
\eea
where we have factorized the term $\lambda_f\equiv V_{f i} V^{*}_{f j}$
(the external fermions are assumed here to be quarks of down type), 
with $V_{ij}$ the corresponding CKM matrix element.

\section{Contribution from the wave-functions renormalization}
\label{Sec.Renorm}
The $\hat T^{\mu\nu}_W$ matrix element corresponding to the vertex corrections 
is ultraviolet divergent and, due to the non-renormalization
theorem of the conserved EMT, it is made finite by adding the contributions 
from the wave-function renormalization on the external legs, namely $\hat T^{\mu\nu}_{CT}$. 
This last contribution can be easily determined by using the following 
method, as illustrated in \cite{Coriano:2012cr}. 
We promote the counterterm SM Lagrangian to a curved background and then extract in the usual way the appropriate renormalized Feynman rules for single insertions of the EMT on the fields of the Standard Model. The metric is taken to be flat after all the functional differentiations. 
Then, for the off-diagonal flavor contributions $(i\neq j)$ to  
$\hat T^{\mu\nu}_{CT}$  we have
\bea
\label{ct}
\hat T^{\mu\nu}_{CT} &=& i \langle p_i, f_i | T^{\mu\nu}_{CT}(0) | p_j, f_j \rangle = 
\frac{i}{4} \bar u_i(p_i) \bigg\{ \left( \gamma^{\mu} p^{\nu}+\gamma^{\nu} p^{\mu}\right) \left( C^{L+}_{ij} P_L + C^{R+}_{ij} P_R \right) \nn \\
&& + \,2 \, \eta^{\mu\nu} \bigg[ C^{L-}_{ij}  \left( m_i P_L - m_j P_R \right) + C^{R-}_{ij}  \left( m_i P_R - m_j P_L \right) \bigg]
\bigg\} u_j(p_j) \,, 
\eea
where
\bea
C^{L \pm}_{ij} = \frac{1}{2} \left( \delta Z^L_{ij} \pm \delta Z^{L \dag}_{ij} \right) \,, \qquad   C^{R \pm}_{ij} = \frac{1}{2} \left( \delta Z^R_{ij} \pm \delta Z^{R \dag}_{ij} \right) \,,
\eea
with $\delta Z^{L,R}_{ij}$ being the fermion wave function renormalization constants. In the on-shell renormalization scheme, which we have chosen for our computation, the renormalization conditions are fixed in terms of the physical parameters of the Standard Model to all orders in the perturbative expansion. In particular for the fermion wave function renormalization constants with $i \ne j$ one obtains
\bea
\delta Z^L_{ij} &=& \frac{2}{m_i^2 - m_j^2} \tRE \bigg\{ m_j^2 \, \Sigma^L_{ij}(m_j^2) + m_i \, m_j \, \Sigma^R_{ij}(m_j^2) + \left( m_i^2 + m_j^2 \right) \Sigma^S_{ij}(m_j^2) \bigg\} \,, \nn \\
\delta Z^R_{ij} &=& \frac{2}{m_i^2 - m_j^2} \tRE \bigg\{ m_j^2 \, \Sigma^R_{ij}(m_j^2) + m_i \, m_j \, \Sigma^L_{ij}(m_j^2) + 2 \, m_i \, m_j \, \Sigma^S_{ij}(m_j^2) \bigg\} \,. 
\eea 
The symbol $\tRE$ gives the real part of the scalar integrals in the self-energies but it has no effect on the CKM matrix elements. Its presence yields $\delta Z^\dag_{ij} = \delta Z_{ij}\left( m_i^2 \leftrightarrow m_j^2 \right)$. Remember also that if the mixing matrix is real $\tRE$ can obviously be replaced with $\rm Re$. \\
For completeness we give the Standard Model flavor changing self-energies ($i \neq j $)
\bea
\label{selfenergies}
\Sigma^L_{ij}(p^2) &=& - \frac{G_F}{4 \pi^2 \sqrt{2}} \sum_f V_{if} V_{fj}^\dag \bigg[ \left( m_f^2 + 2 m_W^2 \right) \mathcal B_1(p^2, m_f^2, m_W^2) + m_W^2 \bigg] \,, \nn \\
\Sigma^R_{ij}(p^2) &=& - \frac{G_F}{4 \pi^2 \sqrt{2}} \, m_i \, m_j \sum_f V_{if} V_{fj}^\dag \, \mathcal B_1(p^2, m_f^2, m_W^2) \,, \nn \\
\Sigma^S_{ij}(p^2) &=& - \frac{G_F}{4 \pi^2 \sqrt{2}} \sum_f V_{if} V_{fj}^\dag \, m_f^2 \, \mathcal B_0(p^2, m_f^2, m_W^2) \,,
\eea
where
\bea
\mathcal B_1(p^2,m_0^2,m_1^2) = \frac{m_1^2 - m_0^2}{2 p^2} \bigg[ \mathcal B_0 (p^2,m_0^2,m_1^2) - \mathcal B_0(0,m_0^2,m_1^2) \bigg] - \frac{1}{2} \mathcal B_0(p^2,m_0^2,m_1^2) \,.
\eea
We have explicitly checked that the counterterm in Eq.(\ref{ct}) is indeed sufficient to remove all the ultraviolet divergences of the $\hat T^{\mu\nu}_W$ matrix element so that $\hat T^{\mu\nu}$ is finite, as expected.
\section{The Ward identity from the conservation of the EMT}
\label{Sec.WardId}
The conservation of the energy-momentum tensor constraints the $\hat T^{\mu\nu}$ matrix element reducing the 12 form factors defined above to a smaller subset of 6 independent contributions. We can derive the Ward identity by imposing the invariance of the 1-particle irreducible generating functional - which depends on the external gravitational metric - under a diffeomorphism transformation and then functional differentiating with respect to the fermion fields. We omit the details of this procedure, which has been discussed extensively in \cite{Coriano:2011zk} and \cite{Coriano:2012cr} for the $TVV'$ and the $Tf\bar{f}$ vertices respectively. The analysis, in this new case, follows similar steps.  
In momentum space, for the unrenormalized matrix element we obtain the Ward identity
\bea
\label{unrenormWI}
q_{\mu} \hat T^{\mu\nu}_W = \bar u_i(p_i) \bigg\{ p_i^\nu \Gamma_{ij}(p_i) - p_j^\nu \Gamma_{ij}(p_j) + \frac{q_\mu}{2} \left( \Gamma_{ij}(p_i) \sigma^{\mu\nu} - \sigma^{\mu\nu} \Gamma_{ij}(p_j) \right)\bigg\} u_j(p_j) \,,
\eea
where $\sigma^{\mu\nu}= [\gamma^{\mu},\gamma^{\nu}]/4$ and $\Gamma_{ij}(p)$ is the fermion two-point function which is given by
\bea
\Gamma_{ij}(p) = i \bigg[ \Sigma_{ij}^L(p^2) \, \psl \, P_L + \Sigma_{ij}^R(p^2) \, \psl \, P_R + \Sigma^S_{ij}(p^2) \left( m_i \, P_L + m_j \, P_R \right) \bigg] \,.
\eea
The off-diagonal (in flavor space) two-point form factors $\Sigma^{L,R,S}(p^2)$ are explicitly given in Eq.(\ref{selfenergies}). \\
The renormalized Ward identity is instead much simpler than Eq.(\ref{unrenormWI}) being just $q_{\mu} \hat T^{\mu\nu} = 0$. It implies a set of homogeneous equations \cite{Degrassi:2008mw} for the renormalized form factors $f_k(p,q)$
\bea
p \cdot q \, f_1(p,q) + q^2 f_2(p,q) &=& 0 \,, \nn \\
f_3(p,q) + q^2 f_5(p,q) + p \cdot q \, f_6(p,q) + \frac{p \cdot q}{2 m_W^2} f_{12}(p,q) &=& 0 \,, \nn \\
p \cdot q \, f_4(p,q) +q^2 f_6(p,q) + \frac{p \cdot q}{2 m_W^2} f_{11}(p,q) &=& 0 \,, \nn \\
f_2(p,q) + f_7(p,q) + q^2 f_9(p,q) + p \cdot q \, f_{10}(p,q) - \frac{p^2 + q^2}{4 m_W^2} f_{12}(p,q) &=& 0 \,, \nn \\
f_1(p,q) + p \cdot q \, f_8(p,q) + q^2 f_{10}(p,q) - \frac{p^2 + q^2}{4 m_W^2} f_{11}(p,q) &=& 0 \,, \nn \\
p \cdot q \, f_{11}(p,q) + q^2 f_{12}(p,q) &=& 0 \,,  
\eea
which provide a strong test on the correctness of our results and allow to reduce the number of independent contributions to the $\hat T^{\mu\nu}$ matrix element.

\section{Flavor-changing form factors}
In this section we present the explicit expressions of the renormalized form factors $F_k$ defined above. They have been computed in the on-shell case retaining the full dependence on the internal ($m_f, m_W$) and external masses ($m_i, m_j$) and on the virtuality, $q^2$, of the graviton line. They are expressed in terms of the dimensionless ratios $x_S = (m_i^2 + m_j^2)/q^2$, $x_D = (m_j^2 - m_i^2)/q^2$, $x_f = m_f^2/q^2$, $x_W = m_W^2/q^2$ and of the combination $\lambda = x_{D}^2-2 x_{S}+1$. We recall that $m_f$ is the mass of the fermion of flavor $f$ running in the loop. \\
Due to their complexity we expand our results onto a basis of massive one-, two- and three-point scalar integrals as
\bea
F_{k}(p,q,m_f) = \sum_{l=0}^{7} C_k^l \, I_l 
\label{Fi}
\eea
where
\bea
\begin{array}{ll}
I_0 = 1 \,, 	& 		I_4 = \mathcal B_0(q^2, m_f^2, m_f^2) \,, \\
I_1 =  \mathcal A_0(m_f^2) - \mathcal A_0(m_W^2) \,,   &   I_5 = B_0(q^2, m_W^2, m_W^2) \,,\\
I_2 =  \mathcal B_0(m_j^2, m_f^2, m_W^2) \,,        &   I_6 = \mathcal C_0(m_j^2, q^2, m_i^2, m_f^2, m_W^2, m_W^2) \,, \\
I_3 = \mathcal B_0(m_i^2, m_f^2, m_W^2) \,,      & I_7 = \mathcal C_0(m_j^2, q^2, m_i^2, m_W^2, m_f^2, m_f^2) \,.
\end{array}
\eea
We give the explicit results for the renormalized form factors $F_1$, $F_3$, $F_4$, $F_7$, $F_8$ and $F_{11}$ while the remaining six can be obtained exploiting the Ward identities derived in the previous section
\bea
\label{WIFF}
F_2 &=& - x_D \, F_1 \,, \nn \\
F_5 &=&  - \frac{1}{q^2} F_3 + x_D^2 \, F_4 + \frac{x_D^2}{m_W^2} F_{11} \,, \nn \\
F_6  &=& - x_D \,  F_4 - \frac{x_D}{2 m_W^2}  F_{11}  \,, \nn \\
F_9  &=& 2 \frac{x_D}{q^2} F_1 - \frac{1}{q^2} F_7 + x_D^2 \, F_8 - \frac{x_S \, x_D}{ m_W^2}  F_{11} \,, \nn \\
F_{10}  &=& - \frac{1}{q^2} F_1 - x_D \, F_8 + \frac{x_S}{2  m_W^2} F_{11}  \,, \nn \\
F_{12} &=& - x_D \, F_{11} \,. 
\eea
The coefficients $C_k^l$ defining $F_k$ in Eq.(\ref{Fi})  are given
in the Appendix \ref{formfactors}.
Finally we remark that the $F_3$, $F_5$, $F_7$ and $F_9$ form factors depend also from the parameter $\chi$ which appears in the gravitational coupling of the $\phi^\pm$ Goldstone bosons through the improved energy-momentum tensor.

\section{Conclusions}
We have presented the computation of the structure of the gravitational form factors of the Standard Model fermions in the  
off-diagonal flavor sector. The analysis has been developed according to our previous study \cite{Coriano:2012cr} where we have discussed the electroweak corrections in the flavor conserving case. The work extends a previous investigation \cite{Degrassi:2008mw} of the same flavor-changing vertex in which the external mass dependence has not been included. The exact expressions presented in our work are relevant for a phenomenological study of the small and intermediate momentum behavior of these form factors, which we plan to address in the near future. 

\vspace{1cm}
\vbox{
\noindent{{\bf Acknowledgements} } \\
\noindent
E.G. would like to thank the PH-TH division of CERN for its kind  hospitality during the preparation of this work. This work was supported by the ESF grant MTT60,  by the recurrent financing SF0690030s09 project and by the European Union through the European Regional Development Fund.
}

\newpage

\appendix

\section{Feynman rules}
\label{feynrules}

We collect here all the Feynman rules involving a graviton that have been used in this work. All the momenta are incoming
\begin{itemize}
\item{ graviton - gauge boson - gauge boson vertex}
\\ \\
\begin{minipage}{95pt}
\includegraphics[scale=1.0]{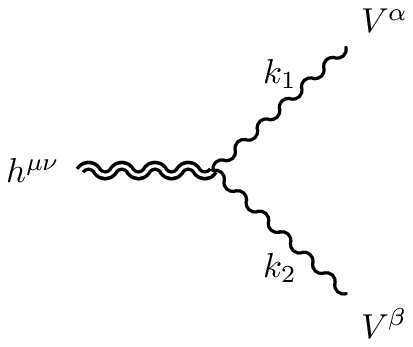}
\end{minipage}
\begin{minipage}{70pt}
\bea
= - i \frac{\kappa}{2} \bigg\{ \left( k_1 \cdot k_2  + M_V^2 \right) C^{\mu\nu\alpha\beta}
+ D^{\mu\nu\alpha\beta}(k_1,k_2) + \frac{1}{\xi}E^{\mu\nu\alpha\beta}(k_1,k_2) \bigg\}
\nn
\eea
\end{minipage}
\bea
\label{FRhVV}
\eea
where $V$ stands for the vector gauge boson $W$.
\item{graviton - fermion - fermion vertex}
\\ \\
\begin{minipage}{95pt}
\includegraphics[scale=1.0]{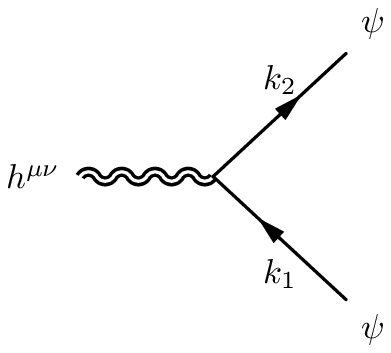}
\end{minipage}
\begin{minipage}{70pt}
\bea
=- i \frac{\kappa}{8} \bigg\{ \gamma^\mu \, (k_1 + k_2)^\nu + \gamma^\nu \,(k_1 + k_2)^\mu - 2 \, \eta^{\mu\nu} \left( \ksl_1 + \ksl_2 - 2 m_f \right)\bigg\}
\nn
\eea
\end{minipage}
\bea
\label{FRhFF}
\eea
\item{graviton - scalar - scalar vertex}
\\ \\
\begin{minipage}{95pt}
\includegraphics[scale=1.0]{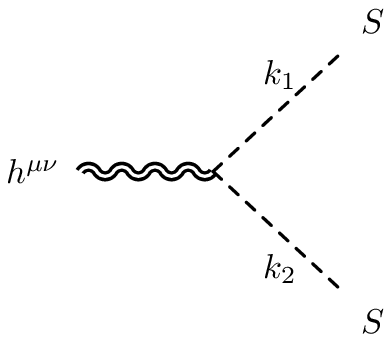}
\end{minipage}
\begin{minipage}{70pt}
\bea
&=&  i \frac{\kappa}{2} \bigg\{ k_{1\, \rho} \, k_{2 \, \sigma} \, C^{\mu\nu\rho\sigma}  - M_S^2 \, \eta^{\mu\nu} \bigg\} \nn \\
&-&  i \frac{\kappa}{2}  2 \chi  \bigg\{ (k_1+k_2)^{\mu}(k_1+k_2)^{\nu} - \eta^{\mu\nu} (k_1+k_2)^2 \bigg\} \nn
\eea
\end{minipage}
\bea
\label{FRhSS}
\eea
where $S$ stands for the Goldstone $\phi^{\pm}$ of the gauge boson $W$. The first line is the contribution coming from the minimal energy-momentum tensor while the second is due to the improvement term.
\item{graviton - scalar - fermion - fermion vertex}
\\ \\
\begin{minipage}{95pt}
\includegraphics[scale=1.0]{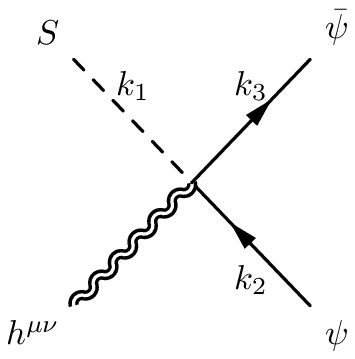}
\end{minipage}
\begin{minipage}{70pt}
\bea
=  \frac{\kappa}{2} \left( C^L_{S\bar \psi \psi} \, P_L + C^R_{S\bar \psi \psi} \, P_R \right) \, \eta^{\mu\nu}
\nn
\eea
\end{minipage}
\bea
\label{FRhSFF}
\eea
where  the coefficients are defined as
\bea
&& C^L_{\phi^+ \bar \psi \psi} = i \frac{e}{\sqrt{2} s_W} \frac{m_{\bar \psi}}{m_W} V_{\bar \psi \psi} \,, \qquad  C^R_{\phi^+ \bar \psi \psi} = - i \frac{e}{\sqrt{2} s_W} \frac{m_{\psi}}{m_W} V_{\bar \psi \psi} \,, \nn \\
&& C^L_{\phi^- \bar \psi \psi} = - i \frac{e}{\sqrt{2} s_W} \frac{m_{\bar \psi}}{m_W} V^*_{\bar \psi \psi} \,, \qquad  C^R_{\phi^- \bar \psi \psi} =  i \frac{e}{\sqrt{2} s_W} \frac{m_{\psi}}{m_W} V^*_{\bar \psi \psi} \,.
\eea
\item{graviton - gauge boson - fermion - fermion vertex}
\\ \\
\begin{minipage}{95pt}
\includegraphics[scale=1.0]{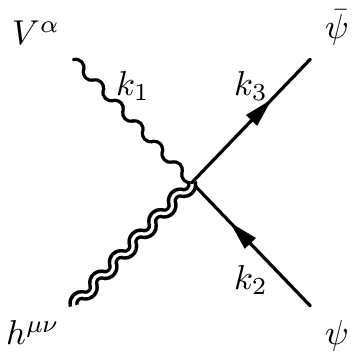}
\end{minipage}
\begin{minipage}{70pt}
\bea
= - \frac{\kappa}{2} \left( C^L_{V \bar\psi \psi} \, P_L + C^R_{V \bar\psi \psi} \, P_R\right) C^{\mu\nu\alpha\beta} \gamma_\beta
\nn
\eea
\end{minipage}
\bea
\label{FRhAFF}
\eea
with
\bea
C^L_{W^+ \bar\psi \psi} = i \frac{e}{\sqrt{2} s_W} V_{\bar\psi \psi}  \,, \quad C^L_{W^- \bar\psi \psi} = i \frac{e}{\sqrt{2} s_W} V^*_{\bar\psi \psi}  \,, \quad C^R_{W^\pm \bar\psi \psi} = 0 \,.
\eea
\end{itemize}
The tensor structures $C$, $D$ and $E$ which appear in the Feynman rules defined above are given by
\bea
&& C_{\mu\nu\rho\sigma} = \eta_{\mu\rho}\, \eta_{\nu\sigma} +\eta_{\mu\sigma} \, \eta_{\nu\rho} -\eta_{\mu\nu} \, \eta_{\rho\sigma} \,, \nn \\
&& D_{\mu\nu\rho\sigma} (k_1, k_2) = \eta_{\mu\nu} \, k_{1 \, \sigma}\, k_{2 \, \rho} - \biggl[\eta^{\mu\sigma} k_1^{\nu} k_2^{\rho} + \eta_{\mu\rho} \, k_{1 \, \sigma} \, k_{2 \, \nu}
  - \eta_{\rho\sigma} \, k_{1 \, \mu} \, k_{2 \, \nu}  + (\mu\leftrightarrow\nu)\biggr] \,, \nn \\
&& E_{\mu\nu\rho\sigma} (k_1, k_2) = \eta_{\mu\nu} \, (k_{1 \, \rho} \, k_{1 \, \sigma} +k_{2 \, \rho} \, k_{2 \, \sigma} +k_{1 \, \rho} \, k_{2 \, \sigma})
-\biggl[\eta_{\nu\sigma} \, k_{1 \, \mu} \, k_{1 \, \rho} +\eta_{\nu\rho} \, k_{2 \, \mu} \, k_{2 \, \sigma} +(\mu\leftrightarrow\nu)\biggr] \,. \nn \\
\eea

\section{Scalar integrals}
\label{scalarint}
In this Appendix we collect the definitions of the scalar integrals appearing in the computation of the matrix element. One-, two- and three- point functions are denoted respectively as $\mathcal A_0$, $\mathcal B_0$ and $\mathcal C_0$ with
\bea
\mathcal A_0(m_0^2) &=& \frac{1}{i \pi^2} \int d^n l \frac{1}{l^2 - m_0^2} \,, \nn \\
\mathcal B_0(p_1^2, m_0^2, m_1^2) &=& \frac{1}{i \pi^2} \int d^n l \frac{1}{(l^2 -m_0^2)((l+p_1)^2 -m_1^2)} \,, \nn \\ 
\mathcal C_0(p_1^2, (p_1-p_2)^2, p_2^2,m_0^2, m_1^2, m_2^2) &=&  \frac{1}{i \pi^2} \int d^n l \frac{1}{(l^2 -m_0^2)((l+p_1)^2 -m_1^2)((l+p_2)^2 -m_2^2)} \,. \nn \\
\eea

\section{Form Factors}
\label{formfactors}
Here we list the coefficients $C_i^j$ which appear in the 
expansion of the form factors
 $F_1$, $F_3$, $F_4$, $F_7$, $F_8$, $F_{11}$ defined in Eq.(\ref{Fi}). 
The remaining ones, as already mentioned, can be computed using the Ward identities in Eq.(\ref{WIFF}). 
\vspace{0.3cm}
\begin{itemize}
\item Coefficients $C_1^i$ entering in $F_1$
\end{itemize}

\small
\bea
C_1^0 &=& 
\frac{q^2}{12 \lambda} \bigg\{ 3 \left(x_D^2 - x_S^2 \right)+6(1-x_S) \left(x_f-6 x_W\right)+16
   \left(x_f-x_W\right) \left(x_f+2 x_W\right) \bigg\} \,, \nn \\
C_1^1 &=& -\frac{2 x_f-3 x_S+4 x_W+3}{6 \lambda} \,, \nn \\
C_1^2 &=& \frac{q^2}{8 \lambda ^2 \, x_D} 
\bigg\{x_D^4 \left(6 x_f+x_S-16 x_W-2\right)+x_D^3 \left(x_S \left(-10 x_f+18  x_W+3\right)+12 \left(x_f x_W+x_f^2  \right. \right. \nn \\
&+& \left. \left. x_f-2 x_W^2\right)-32 x_W-3\right)-x_D^2 \left(x_S^2 \left(20  x_f-46 x_W-6\right)+x_S \left(-28 x_f x_W-4 x_f \left(7 x_f+4\right) \right. \right. \nn \\
&+& \left. \left.  56 x_W^2+54 x_W+4\right)+24 \left(x_f-x_W\right) \left(x_f+2 x_W\right)+6 x_f+x_S^3-26 x_W-1\right) \nn \\
&+&x_D \left(x_S^2 \left(8 x_f+6 x_W+3\right)-2 x_S \left(2 x_f x_W+x_f \left(2 x_f+5\right)-4 x_W^2+x_W\right)-8 \left(x_f-x_W\right) \right. \nn \\
&\times& \left. \left(x_f+2 x_W\right)-3 x_S^3+10 x_W\right)-\left(1-2 x_S\right)^2  \left(2 x_W \left(2 x_f+x_S\right)+\left(x_S-2 x_f\right)^2-8 x_W^2\right)\bigg\} \,, \nn \\
C_1^3 &=& \frac{q^2}{8 \lambda ^2 x_D} 
\bigg\{ x_D^4 \left(-\left(6 x_f+x_S-16 x_W-2\right)\right)+x_D^3 \left(x_S \left(-10 x_f+18 x_W+3\right)+12 \left(x_f x_W+x_f^2 \right.\right. \nn \\
&+& \left.\left. x_f-2 x_W^2\right)-32 x_W-3\right)+x_D^2 \left(x_S^2 \left(20 x_f-46 x_W-6\right)+x_S \left(-28 x_f x_W-4 x_f \left(7 x_f+4\right) \right.\right. \nn \\
&+& \left.\left. 56 x_W^2+54 x_W+4\right)+24  \left(x_f-x_W\right) \left(x_f+2 x_W\right)+6 x_f+x_S^3-26 x_W-1\right) \nn \\
&+& x_D \left(x_S^2 \left(8 x_f+6 x_W+3\right)-2 x_S \left(2 x_f x_W+x_f \left(2 x_f+5\right)-4 x_W^2+x_W\right)-8 \left(x_f-x_W\right) \right. \nn \\ 
&\times& \left. \left(x_f+2 x_W\right)-3 x_S^3+10 x_W\right)+\left(1-2 x_S\right)^2 \left(2 x_W \left(2 x_f+x_S\right)+\left(x_S-2 x_f\right)^2-8 x_W^2\right)\bigg\} \,, \nn \\
C_1^4 &=& \frac{q^2}{12 \lambda ^2} 
\bigg\{x_D^2 \left(3 \left(8 x_f-3\right) x_S-2 x_f \left(16 x_f+32 x_W+11\right)+10 x_W+9\right)-3 x_S^2 \left(8 x_f+26 x_W+3\right) \nn \\ 
&+& 4 x_S \left(-7 x_f x_W+x_f \left(13 x_f+5\right)+42 x_W^2+34 x_W\right)+24 \left(6 x_f-7\right) x_W^2+92 x_f x_W \nn \\
&-& 2 x_f \left(2 x_f+1\right) \left(12 x_f-1\right)+9 x_S^3-96 x_W^3-68 x_W\bigg\} \,, \nn \\
C_1^5 &=& \frac{q^2}{3 \lambda ^2}
\bigg\{ x_f \left(8 x_W \left(x_D^2+x_S-2\right)-2 x_D^2+3 x_S^2-2 x_S-36 x_W^2+1\right) \nn \\
&+& 4 x_W \left(x_D^2 \left(-3 x_S+4 x_W+5\right)+\left(x_S-3 x_W-2\right) \left(3 x_S-2
   x_W-1\right)\right)-12 x_f^2 \left(x_S-1\right)+12 x_f^3\bigg\} \,, \nn \\
C_1^6 &=& \frac{q^4}{2 \lambda ^2} 
\bigg\{ 8 x_W^3 \left(-2 x_D^2+5 x_f+4 x_S-2\right)-4 x_W^2 \left(x_D^2 \left(-2 x_f-2 x_S+3\right)+\left(x_f+x_S\right) \left(6 x_f+x_S\right) \right. \nn \\
&-& \left. 5 x_f-2 x_S\right)-2 x_W \left(-4 x_f^2 \left(x_D^2+x_S-2\right)+3 x_f \left(x_D^2 \left(2 x_S-3\right)-\left(x_S-2\right) x_S\right)+2 \left(x_D^2-x_S^2 \right) \right. \nn \\
&\times& \left. \left(x_D^2-x_S\right) +4 x_f^3\right)+x_f \left(2
   x_f-x_S+1\right) \left(-x_D^2+\left(x_S-2 x_f\right){}^2+4 x_f\right)-16 x_W^4 \bigg\} \,, \nn \\
C_1^7 &=& \frac{q^4}{8 \lambda ^2} 
\bigg\{ 2 x_D^2 \left(x_S \left(x_f \left(-4 x_f+32 x_W+7\right)-8 x_W-2\right)+\left(2-6 x_f\right) x_S^2+2 \left(3-16 x_f\right) x_W^2 \right. \nn \\
&+& \left. 4 x_f \left(4 x_f-9\right) x_W+x_f \left(2 x_f-1\right) \left(8 x_f+5\right)+8 x_W+1\right)+x_D^4 \left(4 x_f-1\right)+16 x_W^3 \left(10 x_f+9 x_S \right. \nn \\
&-& \left.  9\right)-4 x_W^2 \left(34 x_f x_S+24 x_f^2-50 x_f+3 x_S \left(9 x_S-16\right)+24\right)+8 x_W \left(-\left(x_f+8\right) x_S^2+\left(4 \left(x_f \right. \right. \right. \nn \\
&-& \left.\left.\left. 1\right) x_f+6\right) x_S+6 x_f-4 x_f^2 \left(x_f+2\right)+4 x_S^3-2\right)-\left(x_S-2 x_f\right)^2 \left(2 x_f \left(x_S-1\right)-8 x_f^2+3 x_S^2 \right. \nn \\
&-& \left. 4 x_S+2\right)-64 x_W^4\bigg\} \,,
\eea

\normalsize
\vspace{0.3cm}
\begin{itemize}
\item Coefficients $C_3^i$ entering in $F_3$
\end{itemize}

\small
\bea
C_{3}^0 &=& \frac{q^2}{6 \lambda } \bigg\{ -x_S \left(3 x_D^2+32 x_f-20 x_W+3\right)+2 x_f \left(7 x_D^2+4 x_W+9\right)-4 x_W \left(2 x_D^2+4 x_W+3\right) \nn \\
&+& 8 x_f^2+6 x_S^2\bigg\} \,, \nn \\
C_{3}^1 &=& \frac{1}{3 \lambda  \left(x_S^2-x_D^2\right)} \bigg\{ -x_D^2 \left(x_S \left(6 x_f+x_S+12 x_W+4\right)-4 x_f-8 x_W-3\right)+x_D^4 \nn \\
&+& x_S \left(4 x_S-3\right) \left(2 x_f+x_S+4 x_W\right) \bigg\} \,, \nn \\
C_{3}^2 &=& \frac{q^2}{4 \lambda ^2 \left(x_D+x_S\right)} 
\bigg\{ x_D^5 \left(-2 x_f+x_S+2 x_W+2\right)+x_D^4 \left(4 x_W \left(2 x_f+x_S+1\right)-2 x_f x_S+8 x_f^2 \right. \nn \\
&-& \left. 2 x_f-x_S^2+x_S-16 x_W^2\right)+x_D^3 \left(2 x_W \left(2 x_f \left(x_S+3\right)+x_S^2+x_S+2\right)+4 x_f^2 x_S+2 x_f x_S+12 x_f^2 \right. \nn \\
&-& \left. 2 x_f-8 \left(x_S+3\right) x_W^2-x_S^3-2 x_S^2-5 x_S+2\right)+x_D^2 \left(2 x_S^2 \left(-2 x_f+x_W-4\right)+x_S \left(2 x_f \left(2 x_W \right. \right. \right. \nn \\
&+& \left.\left. \left. 9\right)+4 x_f^2-2 x_W \left(4 x_W+9\right)+5\right)-4 x_f x_W-2 x_f \left(2 x_f+5\right)+x_S^3+8 x_W^2+12 x_W-1\right) \nn \\
&+& x_D \left(x_S^3 \left(-8 x_f+4 x_W-1\right)+2 x_S^2 \left(x_f \left(4 x_W+6\right)+4 x_f^2-x_W \left(8 x_W+9\right)+2\right) \right. \nn \\
&-& \left. 2 x_S \left(x_f \left(10 x_W-3\right)+10 x_f^2+\left(3-20 x_W\right) x_W+1\right)+2 \left(5-2 x_f\right) x_W-4 x_f \left(x_f+2\right)+2 x_S^4 \right. \nn \\
&+& \left. 8 x_W^2\right)+x_D^6+x_S^3 \left(-8 x_f+4 x_W+1\right)-2 x_S \left(2 x_f \left(5 x_W+2\right)+10 x_f^2-5 x_W \left(4 x_W+1\right)\right) \nn \\
&+& x_S^2 \left(8 x_f \left(x_W+2\right)+8 x_f^2-2 x_W \left(8 x_W+9\right)-1\right)+4 \left(x_f-x_W\right) \left(x_f+2 x_W\right)+2 x_S^4\bigg\} \nn \\
&+& \chi \frac{2 q^2}{\lambda}  \bigg\{ 2 \left(x_D+1\right) x_f-\left(x_D-1\right) \left(x_D+x_S\right)\bigg\} \,, \nn \\
C_{3}^3 &=& -\frac{q^2}{4 \lambda ^2 \left(x_D-x_S\right)} 
\bigg\{ -x_D^5 \left(-2 x_f+x_S+2 x_W+2\right)+x_D^4 \left(4 x_W \left(2 x_f+x_S+1\right)-2 x_f x_S+8 x_f^2 \right. \nn \\ 
&-& \left. 2 x_f-x_S^2+x_S-16 x_W^2\right)+x_D^3 \left(-2 x_W \left(2 x_f \left(x_S+3\right)+x_S^2+x_S+2\right)-4 x_f^2 x_S-2 x_f x_S-12 x_f^2 \right. \nn \\
&+& \left. 2 x_f+8 \left(x_S+3\right) x_W^2+x_S^3+2 x_S^2+5 x_S-2\right)+x_D^2 \left(2 x_S^2 \left(-2 x_f+x_W-4\right)+x_S \left(2 x_f \left(2 x_W \right.\right.\right. \nn \\ 
&+& \left.\left.\left. 9\right)+4 x_f^2-2 x_W \left(4 x_W+9\right)+5\right)-2 x_f \left(2 x_W+5\right)-4 x_f^2+x_S^3+4 x_W \left(2 x_W+3\right)-1\right) \nn \\
&+& x_D \left(x_S^3 \left(8 x_f-4 x_W+1\right)-2 x_S^2 \left(x_f \left(4 x_W+6\right)+4 x_f^2-x_W \left(8 x_W+9\right)+2\right) \right. \nn \\
&+& \left. x_S \left(x_f \left(20 x_W-6\right)+20 x_f^2-40 x_W^2+6 x_W+2\right)+4 x_f \left(x_W+2\right)+4 x_f^2-2 x_S^4-2 x_W \left(4 x_W \right. \right. \nn \\
&+& \left.\left. 5\right)\right)+x_D^6+x_S^3 \left(-8 x_f+4 x_W+1\right)-2 x_S \left(2 x_f \left(5 x_W+2\right)+10 x_f^2-5 x_W \left(4 x_W+1\right)\right) \nn \\
&+& x_S^2 \left(8 x_f \left(x_W+2\right)+8 x_f^2-2 x_W \left(8 x_W+9\right)-1\right)+4 \left(x_f-x_W\right) \left(x_f+2 x_W\right)+2 x_S^4\bigg\} \nn \\
&-&\chi \frac{2 q^2}{\lambda} \bigg\{ 2 \left(x_D-1\right) x_f+\left(x_D+1\right) \left(x_D-x_S\right)\bigg\} \,, \nn \\
C_{3}^4 &=& \frac{q^2}{6 \lambda ^2} 
\bigg\{ x_D^2 \left(-12 x_S \left(2 x_f+x_W\right)+4 x_f \left(3-8 x_W\right)-28 x_f^2+3 x_S^2+4 x_W \left(3 x_W+5\right)+1\right) \nn \\
&+& x_D^4 \left(8 x_f-2\right)+x_S \left(4 x_f \left(4 x_W-7\right)+44 x_f^2+4 x_W \left(9 x_W-1\right)-1\right)+\left(22 x_f+2\right) x_S^2 \nn \\
&-& 8 \left(-x_f x_W \left(9 x_W+2\right)+3 x_f^3+2 x_f^2+6 x_W^2 \left(x_W+1\right)\right)+10 x_f-3 x_S^3-4 x_W\bigg\} \,, \nn \\
C_{3}^5 &=& -\frac{q^2}{6 \lambda ^2} 
\bigg\{ -x_D^2 \left(4 \left(x_f-10\right) x_W+2 x_f+15 x_S \left(2 x_W+1\right)+44 x_W^2-7\right)+x_D^4 \left(8 x_W+4\right) \nn \\
&+& x_S^2 \left(-18 x_f+28 x_W+5\right)+x_S \left(8 x_f \left(x_W+5\right)+36 x_f^2+26 x_W \left(2 x_W-3\right)-4\right) \nn \\
&-& 4 \left(x_f \left(-18 x_W^2+x_W+5\right)+6 x_f^3+9 x_f^2+2 x_W \left(6 x_W^2+x_W-4\right)\right)+3 x_S^3\bigg\} \nn \\
&+& \chi \frac{4 q^2}{\lambda } \left(x_D^2-2 x_f-x_S\right) \,, \nn \\
C_{3}^6 &=& \frac{q^4}{4 \lambda ^2} 
\bigg\{ 40 x_W^3 \left(-x_D^2+2 x_f+x_S\right)-4 x_W^2 \left(-x_D^2 \left(8 x_f+4 x_S+5\right)+2 x_D^4+4 \left(x_f+2\right) x_S \right. \nn \\
&+& \left. 4 \left(3 x_f^2+x_f-1\right)+3 x_S^2\right)+2 x_W \left(x_D^2-2 x_f-x_S\right) \left(5 x_D^2-4 \left(x_f+3\right) x_S+4 x_f \left(x_f+1\right)+x_S^2 \right. \nn \\
&+& \left. 6\right)+4 x_D^2 x_S-x_D^4-2 x_D^2-8 x_f x_S^3+24 x_f^2 x_S^2+24 x_f x_S^2-32 x_f^3 x_S-48 x_f^2 x_S-24 x_f x_S \nn \\ 
&+& 16 x_f^4+32 x_f^3+24 x_f^2+8 x_f+x_S^4-4 x_S^3+2 x_S^2-32 x_W^4\bigg\} \nn \\
&-& \chi \frac{2 q^4}{\lambda } \bigg\{ x_D^2 \left(2 x_W-1\right)-2 x_W \left(2 x_f+x_S\right)+\left(x_S-2 x_f\right)^2+4 x_f \bigg\} \,, \nn \\
C_{3}^7 &=& -\frac{q^4}{4 \lambda ^2} \bigg\{ -2 x_W \left(x_D^2+2 x_f+x_S-2\right)+\left(-2 x_f+x_S-1\right) \left(x_D^2+2 x_f-x_S\right)+8 x_W^2\bigg\} \nn \\
&\times& \bigg\{ x_D^2 \left(4 x_f-1\right)-4 x_S \left(x_f+x_W\right)+4 \left(\left(x_f-x_W\right)^2+x_W\right)+x_S^2\bigg\} \,,
\eea

\normalsize
\vspace{0.3cm}
\begin{itemize}
\item Coefficients $C_4^i$ entering in $F_4$
\end{itemize}

\small

\bea
C_{4}^0 &=& \frac{2}{3 \lambda ^2} \bigg\{ x_f \left(x_D^2-7 x_S+10 x_W+6\right)-x_W \left(7 x_D^2-19 x_S+20 x_W+12\right)+10 x_f^2\bigg\} \,, \nn \\
C_{4}^1 &=& \frac{2}{3 q^2 \lambda ^2 \left(x_D^2-x_S^2\right)}   
\bigg\{ x_D^2 \left(6 x_f x_S-10 x_f+12 x_S x_W-2 x_S^2+x_S-20 x_W-3\right)+2 x_D^4 \nn \\
&-& \left(x_S-3\right) x_S \left(2 x_f+x_S+4 x_W\right)\bigg\} \,, \nn \\
C_{4}^2 &=& \frac{1}{\lambda ^3 \left(x_D+x_S\right)}  
\bigg\{ x_D^4 \left(x_f \left(x_S+4 x_W+2\right)+4 x_f^2+x_W \left(-9 x_S-8 x_W+4\right)\right)+x_D^3 \left(x_f \left(2 \left(x_S \right. \right. \right. \nn \\
&+& \left. \left. \left. 6\right) x_W+x_S^2-2 x_S+7\right)+2 x_f^2 \left(x_S+6\right)-x_W \left(4 \left(x_S+6\right) x_W+7 x_S^2-10 x_S+5\right)\right) \nn \\
&+& 2 x_D^2 \left(x_f \left(2 x_S \left(-4 x_S+5 x_W+6\right)-7 x_W-4\right)+x_f^2 \left(10 x_S-7\right)+x_W \left(5 x_S \left(3 x_S-4 x_W-4\right) \right. \right. \nn \\
&+& \left.\left. 14 x_W+7\right)\right)+x_D \left(-x_f \left(x_S \left(x_S \left(12 x_S-16 x_W-15\right)+22 x_W+6\right)+8 x_W+3\right) \right. \nn \\
&+& \left. 2 x_f^2 \left(x_S \left(8 x_S-11\right)-4\right)+x_W \left(x_S \left(x_S \left(24 x_S-32 x_W-29\right)+44 x_W+2\right)+16 x_W+7\right)\right) \nn \\
&-& 2 x_D^5 x_W+x_W \left(2 x_f \left(2 \left(x_S-4\right) x_S+1\right)+x_S \left(6 \left(x_S-2\right) x_S+7\right)\right)+x_f \left(x_S \left(4 x_f \left(x_S-4\right) \right. \right. \nn \\
&-& \left.\left. 2 x_S^2+2 x_S-3\right)+2 x_f\right)-4 \left(2 \left(x_S-4\right) x_S+1\right) x_W^2 \bigg\} \,, \nn \\
C_{4}^3 &=& \frac{1}{\lambda ^3 \left(x_D-x_S\right)} 
\bigg\{-x_D^4 \left(x_f \left(x_S+4 x_W+2\right)+4 x_f^2+x_W \left(-9 x_S-8 x_W+4\right)\right)+x_D^3 \left(x_f \left(2 \left(x_S \right. \right. \right. \nn \\
&+&  \left. \left. \left.  6\right) x_W+x_S^2-2 x_S+7\right)+2 x_f^2 \left(x_S+6\right)-x_W \left(4 \left(x_S+6\right) x_W+7 x_S^2-10 x_S+5\right)\right) \nn \\
&+& 2 x_D^2 \left(x_f \left(2 x_S \left(4 x_S-5 x_W-6\right)+7 x_W+4\right)+x_f^2 \left(7-10 x_S\right)+x_W \left(5 x_S \left(-3 x_S+4 x_W+4\right) \right. \right. \nn \\
&-& \left.\left. 7 \left(2 x_W+1\right)\right)\right)+x_D \left(-x_f \left(x_S \left(x_S \left(12 x_S-16 x_W-15\right)+22 x_W+6\right)+8 x_W+3\right) \right. \nn \\
&+& \left. 2 x_f^2 \left(x_S \left(8 x_S-11\right)-4\right)+x_W \left(x_S \left(x_S \left(24 x_S-32 x_W-29\right)+44 x_W+2\right)+16 x_W+7\right)\right) \nn \\
&-& 2 x_D^5 x_W-x_W \left(2 x_f \left(2 \left(x_S-4\right) x_S+1\right)+x_S \left(6 \left(x_S-2\right) x_S+7\right)\right)+x_f \left(x_S \left(2 x_S \left(-2 x_f \right. \right. \right. \nn \\
&+& \left.\left.\left. x_S-1\right)+16 x_f+3\right)-2 x_f\right)+4 \left(2 \left(x_S-4\right) x_S+1\right) x_W^2 \bigg\} \,, \nn \\
C_{4}^4 &=& \frac{1}{6 \lambda ^3} \bigg\{ x_D^2 \left(x_S \left(36 x_f-6 x_W-9\right)-2 \left(2 x_f x_W+5 x_f \left(8 x_f+3\right)+6 x_W^2\right)+6 x_S^2+40 x_W+11\right) \nn \\
&-& 4 x_D^4 \left(2 x_f+1\right)-x_S^2 \left(10 x_f+108 x_W+5\right)+2 x_S \left(2 \left(-58 x_f x_W+25 x_f^2+x_f+81 x_W^2\right) + 77 x_W\right. \nn \\
&-& \left. 1\right)-4 \left(6 \left(13-15 x_f\right) x_W^2-59 x_f x_W+x_f \left(5 x_f \left(6 x_f+1\right)-2\right)+60 x_W^3+20 x_W\right)+3 x_S^3 \bigg\} \,, \nn \\
C_{4}^5 &=& \frac{1}{6 \lambda ^3} 
\bigg\{ -x_D^2 \left(-3 x_S \left(8 x_f-26 x_W+3\right)+50 x_f \left(2 x_W+1\right)+24 x_f^2+6 x_S^2-4 x_W \left(55 x_W+4\right) \right. \nn \\
&+& \left. 11\right)+4 x_D^4 \left(8 x_W+1\right)-20 x_W^2 \left(18 x_f+13 x_S-2\right)+2 \left(2 x_S-1\right) x_W \left(50 x_f+7 x_S+8\right) \nn \\
&+& 42 x_f x_S^2-132 x_f^2 x_S-56 x_f x_S+120 x_f^3+156 x_f^2+40 x_f-3 x_S^3+5 x_S^2+2 x_S+240 x_W^3 \bigg\} \,, \nn \\
C_{4}^6 &=& \frac{q^2}{4 \lambda ^3} 
\bigg\{ x_D^4 \left(2 x_W \left(8 x_f-4 x_S+5\right)+4 x_f-2 x_S-56 x_W^2+3\right)+2 x_D^2 \left(x_S^2 \left(-6 x_f+13 x_W-2\right) \right. \nn \\
&+& \left. x_S \left(-5 \left(4 x_f+5\right) x_W+12 x_f^2+16 x_f+52 x_W^2+1\right)+2 \left(5 \left(12 x_f+1\right) x_W^2+\left(5-6 x_f\right) x_f x_W \right. \right. \nn \\
&-& \left.\left. x_f \left(4 x_f \left(x_f+3\right)+7\right)-50 x_W^3\right)+x_S^3+10 x_W-1\right)-2 x_S^3 \left(8 x_f+5 x_W+1\right)+x_S^2 \left(x_f \left(28 \right.\right. \nn \\
&-& \left.\left. 44 x_W\right)+72 x_f^2+4 \left(6-11 x_W\right) x_W+2\right)-4 x_S \left(6 x_f^2 \left(5-7 x_W\right)+x_f \left(60 x_W^2-26 x_W+4\right) \right. \nn \\
&+& \left. 32 x_f^3+x_W \left(10 \left(1-5 x_W\right) x_W+3\right)\right)+8 \left(50 x_f x_W^3+2 \left(1-15 x_f^2\right) x_W^2-x_f \left(2 x_f \left(5 x_f+9\right) \right. \right. \nn \\
&+& \left.\left. 7\right) x_W+x_f \left(x_f+1\right) \left(2 x_f \left(5 x_f+4\right)+1\right)-20 x_W^4\right)+x_S^4\bigg\} \,, \nn \\
C_{4}^7 &=& \frac{q^2}{4 \lambda ^3} 
\bigg\{x_D^4 \left(4 x_f \left(4 x_f-2 x_W+1\right)+2 x_S-2 x_W-3\right)-2 x_D^2 \left(x_S^2 \left(6 x_f-3 x_W-2\right) \right. \nn \\
&+& \left. x_S \left(10 x_f \left(2 x_f-4 x_W-1\right)+17 x_W+1\right)+6 \left(4 x_f-3\right) x_W^2+2 x_f \left(6 x_f+25\right) x_W-4 x_f \left(10 x_f^2 \right. \right. \nn \\
&+& \left.\left. x_f-2\right)+x_S^3+4 x_W^3-16 x_W-1\right)+8 x_W^3 \left(50 x_f+37 x_S-36\right)-12 x_W^2 \left(8 \left(4 x_f-3\right) x_S \right. \nn \\
&+& \left. 4 \left(x_f \left(5 x_f-9\right)+3\right)+15 x_S^2\right)+2 x_W \left(\left(26 x_f-34\right) x_S^2+\left(28 x_f \left(3 x_f-2\right)+22\right) x_S+44 x_f \right. \nn \\
&-& \left. 8 x_f^2 \left(5 x_f+9\right)+19 x_S^3-8\right)-\left(x_S-2 x_f\right)^2 \left(-20 x_f^2+\left(x_S-2\right) x_S+2\right)-160 x_W^4\bigg\} \,, 
\eea

\normalsize
\vspace{0.3cm}
\begin{itemize}
\item Coefficients $C_7^i$ entering in $F_7$
\end{itemize}

\small

\bea
C_{7}^0 &=& \frac{q^2 x_D}{6 \lambda } \bigg\{ 3 x_D^2+x_S \left(4 x_f-4 x_W-6\right)-8 \left(x_f-x_W\right) \left(x_f+2 x_W\right)-4 x_f+4 x_W+3\bigg\} \,, \nn \\
C_{7}^1 &=& \frac{2 x_D }{3 \lambda  \left(x_D^2-x_S^2\right)} 
\bigg\{ x_D^2 \left(-\left(x_f+x_S+2 x_W-1\right)\right)+x_S \left(x_S \left(-2 x_f+x_S-4 x_W-1\right) \right. \nn \\
&+& \left. 6 \left(x_f+2 x_W\right)\right)-3 \left(x_f+2 x_W\right)\bigg\} \,, \nn \\
C_{7}^2 &=& \frac{q^2}{4 \lambda ^2 \left(x_D+x_S\right)} 
\bigg\{ -x_D^4 \left(x_f \left(-2 x_S+4 x_W+2\right)+4 x_f^2+2 x_W \left(6 x_S-4 x_W+1\right)-1\right) \nn \\
&+& x_D^3 \left(3 x_S^2 \left(4 x_f-4 x_W-1\right)+x_S \left(34 x_W-6 \left(4 x_f x_W+4 x_f^2+x_f-8 x_W^2\right)\right)+16 \left(x_f-x_W\right) \right. \nn \\
&\times& \left. \left(x_f+2 x_W\right)+2 x_f-18 x_W+1\right)+x_D^2 \left(x_S^3 \left(8 x_f-8 x_W-2\right)+x_S^2 \left(-4 x_f \left(4 x_W+1\right)-16 x_f^2 \right. \right. \nn \\
&+& \left.\left. 4 x_W \left(8 x_W+9\right)+1\right)-2 x_S \left(3 x_f+x_W+1\right)+16 \left(x_f-x_W\right) \left(x_f+2 x_W\right)+6 x_f-10 x_W+1\right) \nn \\
&+& x_D \left(x_S \left(x_S \left(-24 x_f+8 x_W-7\right)+8 \left(x_f-x_W\right) \left(x_f+2 x_W\right)+22 x_f+6 x_S^2-2 x_W+2\right) \right. \nn \\
&-& \left. 2 \left(2 x_f+x_W\right)\right)+x_D^5 \left(-2 x_f-8 x_W+1\right)+\left(1-2 x_S\right){}^2 \left(-2 x_f+x_S+2 x_W\right) \left(x_S \right. \nn \\
&-& \left. 2 \left(x_f+2 x_W\right)\right)\bigg\} 
- \chi \frac{2 q^2}{\lambda} \bigg\{ 2 x_f \left(x_D+2 x_S-1\right)-x_D x_S+x_D^2+x_D-2 x_S^2+x_S\bigg\} \,, \nn \\
C_{7}^3 &=& \frac{q^2}{4 \lambda ^2 \left(x_D-x_S\right)} 
\bigg\{ -x_D^4 \left(x_f \left(-2 x_S+4 x_W+2\right)+4 x_f^2+2 x_W \left(6 x_S-4 x_W+1\right)-1\right) \nn \\
&+& x_D^3 \left(3 x_S^2 \left(-4 x_f+4 x_W+1\right)+x_S \left(6 \left(4 x_f x_W+4 x_f^2+x_f-8 x_W^2\right)-34 x_W\right)-16 \left(x_f-x_W\right) \right. \nn \\
&\times& \left. \left(x_f+2 x_W\right)-2 x_f+18 x_W-1\right)+x_D^2 \left(x_S^3 \left(8 x_f-8 x_W-2\right)+x_S^2 \left(-4 x_f \left(4 x_W+1\right)-16 x_f^2 \right. \right. \nn \\
&+& \left. \left. 4 x_W \left(8 x_W+9\right)+1\right)-2 x_S \left(3 x_f+x_W+1\right)+16 \left(x_f-x_W\right) \left(x_f+2 x_W\right)+6 x_f-10 x_W+1\right) \nn \\
&+& x_D \left(2 \left(2 x_f+x_W\right)-x_S \left(x_S \left(-24 x_f+8 x_W-7\right)+8 \left(x_f-x_W\right) \left(x_f+2 x_W\right)+22 x_f+6 x_S^2 \right. \right. \nn \\
&-& \left. \left. 2 x_W+2\right)\right)+x_D^5 \left(2 x_f+8 x_W-1\right)+\left(1-2 x_S\right)^2 \left(-2 x_f+x_S+2 x_W\right) \left(x_S-2 \left(x_f+2 x_W\right)\right)\bigg\} \nn \\
&+& \chi \frac{2 q^2}{\lambda }  \bigg\{ x_D \left(-2 x_f+x_S-1\right)+x_D^2-\left(2 x_S-1\right) \left(x_S-2 x_f\right)\bigg\} \,, \nn \\
C_{7}^4 &=& -\frac{q^2 x_D}{6 \lambda ^2} \left( -2 x_f+x_S-4 x_W-1\right) \bigg\{ x_D^2 \left(8 x_f-2\right)-2 x_S \left(2 x_f+6 x_W+1\right)+12 \left(x_f-x_W\right)^2 \nn \\
&-&4 x_f+3 x_S^2+12 x_W+1\bigg\} \,, \nn \\
C_{7}^5 &=& \frac{q^2 x_D}{6 \lambda ^2} \bigg\{ -48 x_W^3 + 4 x_W^2 \left(-8 x_D^2+18 x_f+7 x_S+1\right)+\left(-2 x_f+x_S-1\right) \left(x_D^2-4 \left(3 x_f+2\right) x_S \right. \nn \\
&+& \left. 4 \left(3 x_f \left(x_f+1\right)+1\right)+3 x_S^2\right)+2 \lambda \, x_W \left(-8 x_f+4 x_S+9\right)\bigg\}
+ \chi \frac{4 q^2 x_D}{\lambda } \left(2 x_f - x_S + 1\right) \,, \nn \\
C_{7}^6 &=& -\frac{q^4 x_D}{4 \lambda ^2} 
\bigg\{ 2 x_D^2 \left(-x_S \left(x_W \left(8 x_f+4 x_W+23\right)-2\right)+x_W \left(x_f \left(8 x_W-2\right)+8 x_f^2 \right. \right. \nn \\
&+& \left. \left. 2 \left(7-8 x_W\right) x_W+9\right)+2 x_S^2 x_W-1\right)+x_D^4 \left(12 x_W-1\right)-2 x_S^3 \left(4 x_f+3 x_W+2\right) \nn \\
&+& x_S^2 \left(4 \left(5 x_f x_W+6 x_f \left(x_f+1\right)+x_W^2\right)+42 x_W+2\right)-4 x_S \left(2 x_f^2 \left(x_W+6\right)+x_f \left(6-4 x_W \left(x_W \right. \right. \right. \nn \\
&+& \left.\left.\left. 1\right)\right)+8 x_f^3+x_W \left(2 \left(5-3 x_W\right) x_W+7\right)\right)+8 \left(\left(10 x_f+1\right) x_W^3-2 \left(x_f+1\right) \left(3 x_f-1\right) x_W^2 \right. \nn \\
&-& \left. x_f \left(2 x_f^2+x_f+2\right) x_W+x_f \left(x_f+1\right) \left(2 x_f \left(x_f+1\right)+1\right)-4 x_W^4\right)+x_S^4+4 x_W\bigg\} \nn \\
&+& \chi \frac{2 q^4 x_D}{\lambda }  \bigg\{ -x_D^2+\left(x_S-2 x_f\right) \left(-2 x_f+x_S+2 x_W\right)+4 x_f-2 x_W\bigg\} \,, \nn \\
C_{7}^7 &=& \frac{q^4 x_D}{4 \lambda ^2}  \left( -2 x_f+x_S-4 x_W-1\right) \left( 2 x_f+x_S-2 x_W-1\right) \bigg\{ x_D^2 \left(4 x_f-1\right)-4 x_S \left(x_f+x_W\right) \nn \\
&+& 4 \left(\left(x_f-x_W\right)^2+x_W\right)+x_S^2\bigg\} \,, 
\eea
\normalsize
\vspace{0.3cm}
\begin{itemize}
\item Coefficients $C_8^i$ entering in $F_8$
\end{itemize}

\small

\bea
C_{8}^0 &=& \frac{10 x_D}{3 \lambda ^2} \left(x_f-x_W\right) \left(-2 x_f+x_S-4 x_W-1\right) \,, \nn \\
C_{8}^1 &=& \frac{2 x_D}{3 q^2 \lambda ^2 \left(x_D^2-x_S^2\right)} \bigg\{ x_D^2 \left(4 x_f-5 x_S+8 x_W+5\right)+x_S \left(5 x_S \left(-2 x_f+x_S-4 x_W-1\right) \right. \nn \\
&+& \left.12 \left(x_f+2 x_W\right)\right)-6 \left(x_f+2 x_W\right)\bigg\} \,, \nn \\
C_{8}^2 &=& \frac{x_W-x_f}{\lambda ^3 \left(x_D+x_S\right)}  
\bigg\{ x_D^4 \left(4 x_f-x_S+8 x_W+6\right)+x_D^3 \left(9 x_S \left(2 x_f-x_S+4 x_W+2\right)-16 x_f-32 x_W \right. \nn \\
&-& \left. 7\right)-2 x_D^2 \left(-2 x_S^2 \left(3 x_f+6 x_W+2\right)+9 x_f+3 x_S^3+x_S+18 x_W+2\right)+x_D \left(x_S \left(x_S \left(24 x_f \right. \right. \right. \nn \\
&-&  \left.\left.\left. 12 x_S+48 x_W+17\right)-10 \left(3 x_f+6 x_W+1\right)\right)+4 x_f+8 x_W+1\right)+2 x_D^5 \nn \\
&-&\left(2 x_S-1\right)^3 \left(x_S-2 \left(x_f+2 x_W\right)\right)\bigg\} \,, \nn \\
C_{8}^3 &=& \frac{x_f-x_W }{\lambda ^3 \left(x_D-x_S\right)}  
\bigg\{ x_D^4 \left(-4 x_f+x_S-8 x_W-6\right)+x_D^3 \left(9 x_S \left(2 x_f-x_S+4 x_W+2\right)-16 x_f-32 x_W \right. \nn \\
&-& \left. 7\right)+2 x_D^2 \left(-2 x_S^2 \left(3 x_f+6 x_W+2\right)+9 x_f+3 x_S^3+x_S+18 x_W+2\right)+x_D \left(x_S \left(x_S \left(24 x_f \right. \right. \right. \nn \\
&-& \left.\left.\left. 12 x_S+48 x_W+17\right)-10 \left(3 x_f+6 x_W+1\right)\right)+4 x_f+8 x_W+1\right)+2 x_D^5 \nn \\
&+& \left(2 x_S-1\right)^3 \left(x_S-2 \left(x_f+2 x_W\right)\right)\bigg\} \,, \nn \\
C_{8}^4 &=& \frac{x_D}{6 \lambda ^3} 
\bigg\{x_D^2 \left(\left(13-16 x_f\right) x_S+32 x_f \left(x_f+2 x_W\right)+26 x_f-34 x_W-13\right)+x_S^2 \left(2 x_f+120 x_W+19\right) \nn \\
&-& 2 x_S \left(-56 x_f x_W+2 x_f \left(x_f+2\right)+150 x_W^2+86 x_W+3\right)+2 \left(-2 x_f \left(90 x_W^2+44 x_W+1\right) \right. \nn \\
&+& \left. 60 x_f^3-14 x_f^2+x_W \left(30 x_W \left(4 x_W+5\right)+43\right)+1\right)-15 x_S^3\bigg\} \,, \nn \\
C_{8}^5 &=& \frac{x_D}{6 \lambda ^3} 
\bigg\{ 4 x_W^2 \left(-16 x_D^2+90 x_f-13 x_S+29\right)+\left(-2 x_f+x_S-1\right) \left(-13 x_D^2-4 \left(15 x_f+1\right) x_S \right. \nn \\
&+& \left. 60 x_f \left(x_f+1\right)+15 x_S^2+2\right)+2 \lambda  x_W \left(-16 x_f+8 x_S-9\right)-240 x_W^3\bigg\} \,, \nn \\
C_{8}^6 &=& \frac{q^2 x_D}{4 \lambda ^3}  
\bigg\{ 2 x_D^2 \left(x_S \left(12 \left(2 x_f x_W-x_f+x_W^2\right)+7 x_W-4\right)-6 \left(4 x_f+3\right) x_W^2-6 x_f \left(4 x_f+5\right) x_W \right. \nn \\
&+& \left. 12 x_f \left(x_f+1\right)+x_S^2 \left(3-6 x_W\right)+48 x_W^3-5 x_W+2\right)+x_D^4 \left(4 x_W-1\right)+2 x_S^3 \left(20 x_f+7 x_W+4\right) \nn \\
&-& 2 x_S^2 \left(18 x_f \left(x_W+2\right)+60 x_f^2+\left(13-6 x_W\right) x_W+2\right)+4 x_S \left(-6 x_f^2 \left(x_W-8\right)+12 x_f \left(-3 x_W^2 \right. \right. \nn \\
&+& \left. \left. x_W+1\right)+40 x_f^3+x_W \left(2 \left(x_W-3\right) x_W+5\right)\right)+4 \left(-2 \left(50 x_f+13\right) x_W^3+6 \left(2 x_f \left(5 x_f+4\right)  \right. \right. \nn \\
&+& \left. \left.  1\right) x_W^2+\left(2 x_f^2 \left(10 x_f+9\right)-1\right) x_W-4 x_f \left(x_f+1\right) \left(5 x_f \left(x_f+1\right)+1\right)+40 x_W^4\right)-5 x_S^4\bigg\} \,, \nn \\
C_{8}^7 &=& \frac{q^2 x_D}{4 \lambda ^3} 
\bigg\{ 2 x_D^2 \left(x_S \left(-2 x_f \left(18 x_W+5\right)+15 x_W+4\right)+\left(6 x_f-3\right) x_S^2+6 \left(8 x_f-3\right) x_W^2 \right. \nn \\
&-&\left.  3 \left(2 x_f \left(4 x_f-9\right)+5\right) x_W-24 x_f^3+8 x_f-2\right)+x_D^4 \left(1-4 x_f\right)-2 x_S^3 \left(2 x_f+25 x_W+4\right) \nn \\
&+& 4 x_S \left(72 \left(x_f-1\right) x_W^2-3 \left(2 \left(x_f-3\right) x_f+5\right) x_W+4 x_f^3-2 x_f-70 x_W^3\right)+x_S^2 \left(x_f \left(8-36 x_W\right) \right. \nn \\
&+&\left. 90 x_W \left(2 x_W+1\right)+4\right)+8 \left(5 \left(7-10 x_f\right) x_W^3+6 \left(x_f-1\right) \left(5 x_f-3\right) x_W^2 \right. \nn \\
&+& \left. x_f \left(2 x_f+3\right) \left(5 x_f-3\right) x_W+2 \left(2-5 x_f\right) x_f^3+20 x_W^4\right)+5 x_S^4+20 x_W\bigg\} \,, 
\eea
\normalsize
\vspace{0.3cm}
\begin{itemize}
\item Coefficients $C_{11}^i$ entering in $F_{11}$
\end{itemize}

\small

\bea
C_{11}^0 &=& \frac{q^2 x_W}{6 \lambda }  \bigg\{ 2 x_f-3 x_S+28 x_W+3 \bigg\} \,, \nn \\
C_{11}^1 &=& \frac{x_W}{3 \lambda  \left( x_D^2-x_S^2\right) } \bigg\{x_S^2 -x_D^2 + 6 \left(1- x_S \right) \left(x_f+2 x_W\right) \bigg\} \,, \nn \\
C_{11}^2 &=& \frac{q^2 x_W}{4 \lambda ^2 x_D \left(x_D+x_S\right)} 
\bigg\{ x_D^4 \left(-2 x_f+x_S+10 x_W+2\right)-x_D^3 \left(x_S \left(6 x_f-40 x_W-5\right)+8 \left(x_f-x_W\right) \right. \nn \\ 
&\times& \left.  \left(x_f+2 x_W\right)+x_S^2+28 x_W+2\right)-x_D^2 \left(2 x_W \left(2 x_f \left(x_S+4\right)-15 x_S^2+11 x_S+8\right)+4 x_f x_S^2 \right. \nn \\
&+& \left. 4 x_f^2 x_S-8 x_f x_S+16  x_f^2+6 x_f-8 \left(x_S+4\right) x_W^2+x_S^3+1\right)+x_D \left(x_S^2 \left(24 x_f-2 x_W+6\right) \right. \nn \\
&-& \left. 2 x_S \left(10 x_f x_W+x_f \left(10 x_f+11\right)-20 x_W^2+4 x_W+1\right)+4 \left(x_f x_W+x_f^2+x_f-2 x_W^2\right) \right. \nn \\
&-& \left. 7 x_S^3-2 x_W\right)+x_D^5-\left(1-2 x_S\right)^2 \left(2 x_W \left(2 x_f+x_S\right)+\left(x_S-2 x_f\right)^2-8 x_W^2\right)\bigg\} \,, \nn \\
C_{11}^3 &=& -\frac{q^2 x_W}{4 \lambda ^2 x_D \left(x_D-x_S\right)} 
\bigg\{ -x_D^4 \left(-2 x_f+x_S+10 x_W+2\right)-x_D^3 \left(x_S \left(6 x_f-40 x_W-5\right) \right. \nn \\
&+& \left. 8 \left(x_f-x_W\right) \left(x_f+2 x_W\right)+x_S^2+28 x_W+2\right)+x_D^2 \left(2 x_W \left(2 x_f \left(x_S+4\right)-15 x_S^2+11 x_S+8\right) \right. \nn \\
&+& \left. 4 x_f x_S^2+4 x_f^2 x_S-8 x_f x_S+16 x_f^2+6 x_f-8 \left(x_S+4\right) x_W^2+x_S^3+1\right)+x_D \left(x_S^2 \left(24 x_f-2 x_W \right. \right. \nn \\
&+& \left. \left. 6\right)-2 x_S \left(10 x_f x_W+x_f \left(10 x_f+11\right)-20 x_W^2+4 x_W+1\right)+4 \left(x_f x_W+x_f^2+x_f-2 x_W^2\right) \right. \nn \\
&-& \left. 7 x_S^3-2 x_W\right)+x_D^5+\left(1-2 x_S\right)^2 \left(2 x_W \left(2 x_f+x_S\right)+\left(x_S-2 x_f\right)^2-8 x_W^2\right)\bigg\} \,, \nn \\
C_{11}^4 &=& \frac{q^2 x_W}{6 \lambda ^2}  \bigg\{ x_D^2 \left(8 x_f-2\right)-2 x_S \left(2 x_f+15 x_W+1\right)+12 \left(x_f-4 x_W\right) \left(x_f-x_W\right) \nn \\
&-& 4 x_f+3 x_S^2+30 x_W+1\bigg\} \,, \nn \\
C_{11}^5 &=& \frac{q^2 x_W}{3 \lambda ^2} \bigg\{ -2 x_D^2 \left(8 x_W+1\right)-2 x_S \left(6 x_f-4 x_W+1\right)+12 \left(4 x_f x_W+x_f^2+x_f-5 x_W^2\right) \nn \\
&+& 3 x_S^2+8 x_W+1\bigg\} \,, \nn \\
C_{11}^6 &=& \frac{q^4 x_W}{2 \lambda ^2} 
\bigg\{ -12 x_W^2 \left(-2 x_D^2+6 x_f+x_S+1\right)-2 x_W \left(x_D^2 \left(4 x_f-2 x_S+3\right)+\left(4 x_f-2\right) x_S \right. \nn \\
&-& \left. 4 x_f \left(3 x_f+2\right)+x_S^2\right)+\left(-2 x_f+x_S-1\right) \left(x_D^2-\left(x_S-2 x_f\right)^2-4 x_f\right)+40 x_W^3\bigg\} \,, \nn \\
C_{11}^7 &=& \frac{q^4 x_W}{4 \lambda ^2} 
\bigg\{ 4 x_W \left(x_D^2 \left(4 x_f-1\right)+4 x_f \left(3 x_f+x_S\right)-8 x_f+x_S \left(3 x_S-4\right)+2\right)-\left(2 x_f+x_S-1\right) \nn \\
&\times& \left(x_D^2 \left(4 x_f-1\right)+\left(x_S-2 x_f\right)^2\right)-36 x_W^2 \left(2 x_f+x_S-1\right)+32 x_W^3\bigg\} \,.
\eea

\end{document}